\documentclass[showpacs,amsmath,amssymb,twocolumn]{revtex4}
\usepackage{graphics}
\usepackage[dvips]{graphicx}

\newcommand{\subtext}[1]{{\mbox{\scriptsize #1}}}
\newcommand{\vare}{\varepsilon}
\newcommand{\lo}{\hbar\omega_\subtext{LO}}
\newcommand{\D}{\displaystyle}

\begin{document}

\title{First order non-equilibrium phase transition and bistability of an electron gas}

\author{M.A.\ Rodr\'\i guez-Meza}
\address{Instituto Nacional de Investigaciones Nucleares, 
  A.P.\ 18--1027 M\'exico D.F. 11801, M\'exico. \\
  e-mail: marioalberto.rodriguez@inin.gob.mx
  }

\date{\today}

\begin{abstract}
We study the carrier concentration bistabilities that occur to a highly 
photo-excited electron gas.
The kinetics of this non-equilibrium electron gas is given by a set of 
nonlinear rate equations. 
For low temperatures and cw photo-excitation we show that they
have three steady state solutions when the photo-excitation energy is in a
certain interval which depends on the electron-electron interaction. 
Two of them are stable and the other is unstable. 
We also find the hysteresis region in terms of which these bistabilities are
expressed.
A diffusion model is constructed which allows the coexistence of two homogeneous
spatially separated phases in the
non-equilibrium electron gas.
The order parameter is the difference of the electron population in the
bottom of the conduction band of these two steady stable states.
By defining a generalized free potential we obtain the Maxwell construction 
that determines the order parameter. 
This order parameter goes to zero when we approach to the critical curve.
Hence, this phase transition is a non-equilibrium first order phase 
transition.

\vspace{0.25\baselineskip}

\vspace{0.75\baselineskip}

Estudiamos las biestabilidades en la concentraci\'on de portadores que
le ocurren a un gas de electrones altamente fotoexcitado.
La cin\'etica de este gas de electrones fuera de equilibrio est\'a dada por
un conjunto de ecuaciones de raz\'on no-lineales.
Para bajas temperaturas y fotoexcitaci\'on en modo continuo (cw) mostramos
que estas ecuaciones tienen tres soluciones de estado estacionario
cuando la energ\'\i a de fotoexcitaci\'on est\'a en un cierto intervalo, el cual
depende de la interacci\'on electr\'on-electr\'on. 
Dos de ellas son estables y la otra es inestable. Tambi\'en, encontramos
la regi\'on de histeresis en t\'erminos de la cual estas biestabilidades son
expresadas.
Construimos un modelo de difusi\'on que permite la coexistencia de dos
fases homog\'eneas espacialmente separadas del gas de electrones
fuera de equilibrio.
El par\'ametro de orden es la diferencia de la poblaci\'on electr\'onica en
el fondo de la banda de conducci\'on de estos dos estados estacionarios estables.
Definiendo un potencial libre generalizado del sistema obtenemos la 
construcci\'on de Maxwell que determina entonces al par\'ametro de orden.
El par\'ametro de orden va a cero cuando nos aproximamos a la curva cr\'\i tica.
Por eso, esta transici\'on de fase es una transici\'on de fase fuera de equilibrio de primer orden.

\vspace{0.25\baselineskip}
\noindent {\em Descriptores}: transici\'on de fase fuera de equilibrio; biestabilidad;
ecuaciones de raz\'on no-lineales; gas de electrones en semiconductores.

\end{abstract}

%\pacs{72.20.Jv}

\maketitle

\section{Introduction}

Bistability, threshold switching transitions, spatial pattern formation, 
self-sustained oscillations and chaos in semiconductors are related to 
non-equilibrium phase transitions\cite{scholl2007,scholl}. These non-equilibrium phase 
transitions in 
semiconductors have been studied in the past decades mainly in connection with 
the 
nonlinear generation-recombination mechanism, including impact 
ionization\cite{scholl,landsberg76,pimpale}. 
Optical and transport properties of semiconductors strongly depend on the 
electron 
population in the bottom of the conduction band and many experimental and 
theoretical 
studies on hot photo excited electron systems have shown that this electron 
population depends on the excitation energy\cite{hotel,shah,carrillo}. 
In this paper we present a 
study 
of 
the carrier concentration bistability 
that occur to an electron gas generated by a laser excitation and 
a study of
a first order non-equilibrium 
phase 
transition between two homogeneous stable steady states of the excited 
electron 
gas. They both
appear when the energy of the pump is varied. The order parameter is the 
difference between the electron populations in the bottom of the conduction 
band 
of the stable states. It depends upon the effectiveness of the 
electron-electron 
interaction. We obtain the order parameter by defining a "potential" which 
allows 
us to make a construction similar to the Maxwell construction for the 
equilibrium 
phase transition of a van der Waals gas\cite{schlogl,mar2001}.

In following section we present the theoretical model we use to study 
a non-equilibrium electron gas in semiconductors. In section third we
deduce the stationary state solutions to the non-linear rate equations
of the model and we make the estability analysis. By considering a
standard first order phase transition of the classical van der Waals gas,
we analyze
the non-equilibrium phase transition of the electron gas and obtain
the corresponding Maxwell construction. We draw our conclusion in
the final section.

\section{Theoretical model of the non-equilibrium electron gas}

We describe the non-equilibrium electron gas in semiconductors using the 
model equations given in Ref.\ \cite{mar}. Here, we give a brief description
of these model equations and refer the reader to the details of their
derivation to Refs.\ \cite{mar2001,mar}.
Electrons in the bottom of the conduction band of a semiconductor play an
important role in the dynamics of the whole conduction electron gas. In
general, electrons with an energy in excess less than the longitudinal
optical (LO) phonon energy can not make transitions by emitting LO phonons.
In the case in which the emission of LO phonons is one of the dominant
mechanisms, the nonequilibrium kinetics of the electron gas in the
conduction band of a semiconductor is given as follows\cite{mar}. 
We define a set
of energy levels, each one of them representing an energy interval of width 
$\Delta \epsilon$ of the conduction band. Although not strictly necessary,
for simplicity $\Delta \epsilon $ is set equal to 
$\hbar \omega_\subtext{LO}$, the
LO phonon energy\cite{mar}. We set the electron population $\chi_i$ in these 
energy levels
and, based on the main interaction mechanisms, the nonlinear rate equations
that give the temporal behavior of these populations are obtained.

When an electron in level $i$ emits a LO phonon it losses an amount of energy 
$\hbar \omega_\subtext{LO}$ and passes to the level $i-1$.
If the frequency of this event is $\nu_{o}^{+}$ then we have that the rate of 
change of population $\chi_i$ is $\nu _{o}^{+}(\chi _{i+1}-\chi _{i})$. Similarly for
the absorpion of a LO phonon we have the rate 
$\nu_{o}^{-}(\chi _{i-1}-\chi _{i})$ where $\nu_{o}^{-}$ is the frequency of
this process.
The electron-electron interaction 
gives
to the rate of change of the
population $\chi _{i}$ at level $i$ 
two terms,
$ZN_{\subtext{max} }\chi_\subtext{tot}(\chi _{i+1}-2\chi _{i}+\chi _{i-1})$ 
and $ZN_{\subtext{max} }\chi_{0}(\chi _{i}-\chi _{i-1})$. 
Their form
come from considering the contribution to the rate of change of the
population $\chi_{i}$ of the interaction between electron populations of
all the energy levels, 
and from the use of energy conservation\cite{mar}.
When an electron suffers a recombination it returns to the valence band and
this gives us the term $-w\chi_i$ with $w$ the frequency of the process.

Then, by collecting all these contributions, we have the following set of rate 
equations which describes the
kinetics of a photo-excited electron gas in semiconductors\cite{mar}, 
\begin{eqnarray}
\frac{d\chi _{i}}{dt} &=&\nu _{o}^{+}(\chi _{i+1}-\chi _{i})+\nu
_{o}^{-}(\chi _{i-1}-\chi _{i}) 
\nonumber \\ &&
+ZN_{\subtext{max} }\chi_\subtext{tot}(\chi _{i+1}-2\chi _{i}+\chi _{i-1}) 
\nonumber \\ &&
+ZN_{\subtext{max} }\chi_{0}(\chi _{i}-\chi _{i-1})
+g_{p}\delta _{i,i_{p}}-w\chi _{i}\quad ,		\label{phasetr_eq_01}
\end{eqnarray}
for $i\neq 0$. For $i=0$, since the emission of LO phonons by an electron is
not possible, we have 
\begin{eqnarray}
\frac{d\chi _{0}}{dt} &=&\nu _{o}^{+}\chi _{1}-\nu _{o}^{-}\chi _{0} 
+ZN_{\subtext{max} }\chi_\subtext{tot}(\chi_{1}-\chi _{0})
\nonumber   \\
&&+ZN_{\subtext{max}}\chi_{0}\chi_{0}+g_{p}\delta_{0,i_{p}}-w\chi_{0}\quad .
\label{phasetr_eq_02}
\end{eqnarray}
The
electron populations $\chi_{i}$ have been normalized to the maximum
reachable electron concentration $N_{\subtext{max} }$ and 
$\chi_\subtext{tot}=\sum \chi_{i}$. 
The terms with the Kronecker delta is the generation contribution with
generation rate $g_p$.
The main interaction mechanisms,
the generation and recombination terms 
depend on the lattice temperature, carrier concentration and
material parameters. 

Due to the electron-electron interaction, this set of rate
equations is non linear. We should say that Eqs.\ (1) and (2) which
are the main equations of our model came from a more general theory
published in Ref.\ \cite{mar}. See, also 
Refs.\ \cite{carrillo,mar2001,lilia} 
for other details of the
model equations and other applications.

We make the following definitions
$\nu \equiv \nu_\subtext{o}^+/ZN_\subtext{max}$, 
$\mu \equiv \nu_\subtext{o}^-/ZN_\subtext{max}$, 
$\omega \equiv w/ZN_\subtext{max}$, 
$\chi_p \equiv g_p/ZN_\subtext{max}$,
and $\vare \equiv \epsilon/\Delta\epsilon$. 
By this normalization, 
we have 
eliminated one parameter. For low temperatures\cite{mar}, 
$\mu\ll\nu$. Also, in steady state, $\chi_p = \omega$,
$\chi_\subtext{tot}=1$ and the total carrier concentration $N=N_\subtext{max}$. 
Hence, 
in this case, we have only two relevant control parameters, $\nu$ and 
$\vare_p$. 
Here $\varepsilon_p=\epsilon_p/\Delta \epsilon$, 
the excitation energy in units of $\Delta$$\epsilon$.

With these definitions and assumptions and considering the
finite differences as derivatives:
\begin{eqnarray*}
\chi_{i+1}-\chi_{i} &\rightarrow &\frac{d\chi }{d\epsilon }\Delta \epsilon 
\\
\chi_{i+1}-2\chi_{i}+\chi_{i-1} &\rightarrow &\frac{d^{2}\chi }{d\epsilon
^{2}}(\Delta \epsilon )^{2} \\
\chi_{i}-\chi_{i-1} &\rightarrow &\frac{d\chi }{d\epsilon }\Delta \epsilon
-\frac{d^{2}\chi }{d\epsilon ^{2}}(\Delta \epsilon )^{2}
\end{eqnarray*}
we have that the equations (\ref{phasetr_eq_01}) and (\ref{phasetr_eq_02}) 
that
give the steady state of a cw photo-excited electron gas in semiconductors
are translated to the following differential equation 
\begin{equation}
0=(1-\chi_{0})\frac{d ^{2}\chi }{d\varepsilon ^{2}}+(\nu +\chi_{0})\frac{
d \chi }{d\varepsilon }+\omega\delta (\varepsilon
-\varepsilon_{p})-\omega \chi   \label{phasetr_eq_03}
\end{equation}

The electron distribution function $\chi (\varepsilon ,\tau )$ is continuous
and positive definite in the whole interval $0\leq \varepsilon <\infty$ and
has a discontinuity in its first derivative at 
$\varepsilon =\varepsilon _{p}$. 
In addition, it must satisfy the conditions 
\begin{eqnarray}
\chi_\subtext{tot} &=&\int_{0}^{\infty }d\varepsilon 
\;\chi (\varepsilon ,\tau )
\label{phasetr_eq_04} \\
\chi _{0} &=&\int_{0}^{1}d\varepsilon \;\chi (\varepsilon ,\tau )
\end{eqnarray}
The Eq. (\ref{phasetr_eq_03}) has the steady state solution\cite{carrillo} 
\[
\chi ^{s}(\varepsilon )=\left\{ 
\begin{array}{lll}
Ae^{\alpha \varepsilon }+Be^{-\beta \varepsilon } & ; & 0\leq \varepsilon
\leq \varepsilon _{p} \\ 
Ce^{-\beta \varepsilon } & ; & \varepsilon _{p}\leq \varepsilon 
\end{array}
\right. 
\]
where 
\begin{equation}
\left. 
\begin{array}{l}
\alpha  \\ 
\beta 
\end{array}
\right\} =\frac{\left[ (\nu +\chi _{0}^{s})^{2}+4\omega (1-\chi
_{0}^{s} )\right] ^{1/2}\mp (\nu +\chi _{0}^{s})}{2(1-\chi
_{0}^{s} )}
\end{equation}
The electron population
at the lowest level $\chi _{0}^{s}$ is determined self-consistently by the
equation 
\begin{equation}
\chi _{0}^{s}=\frac{\omega }{\alpha }e^{-\alpha \varepsilon _{p}}
\frac{e^{\alpha }-e^{-\beta }}{\alpha +\beta }\frac{1}{1-\chi _{0}^{s}}
\label{phasetr_eq_08}
\end{equation}
The coefficients $A$, $B$, and $C$ are obtained using the continuity in 
$\chi (\varepsilon )$, the discontinuity in its first derivative at 
$\varepsilon =\varepsilon _{p}$, and Eq.\ (\ref{phasetr_eq_04}). In
particular, we obtain that $A/B=\alpha /\beta $ (See Ref.\ \cite{carrillo}
for more details).

To conclude this section we give the explicit expression for the 
control parameter $\nu$ that it is needed for the following discussion and
we refer the reader to Refs.\ \cite{mar2001,mar,mar2002} for more details and the 
expressions of
other interaction mechanisms. Then, for $\nu$ we
have\cite{mar2001}
\begin{equation}\label{nu}
  \nu =
    \frac{ 2^{3/2} \lo }{ \sqrt{\pi k_\subtext{B}T_\subtext{e} } }
    \left (	1 - \frac{ {\cal E}_\infty }{ {\cal E}_\subtext{s} }
    \right )
    \left ( N_q + 1 \right ) 
	S_\subtext{LO} S_\subtext{ee}^{-1} A
\end{equation}
where $T_\subtext{e}$ is the effective electron temperature, $k_\subtext{B}$ 
is the Boltzmann constant, ${\cal E}_\infty$ and ${\cal E}_\subtext{s}$ are 
the static and optical dielectric constants, respectively, and $\hbar$ is the
Planck constant. Phonon population effects may be taken into account in $N_q$
which is the phonon population at wavevector of magnitude $q$.
The screening in the electron-LO phonon interaction is taken into account in
the factor\cite{Yoff}
\begin{displaymath}
  S_\subtext{LO} \,=\,\, 
      \left (
1 + \left ( \frac{ N }{ N_\subtext{LO} } 
		\right )^2
      \right )^{-1}
\end{displaymath}
where 
\begin{displaymath}
N_\subtext{LO} = \frac{{\cal E}_\infty m (\lo)^3}
{3^{3/2} 8 \pi e^2 \hbar^2 k_\subtext{B} T_\subtext{e}}
\end{displaymath}
is
the threshold value for the concentration in the conduction band 
at which the screening becomes important\cite{Yoff}. 
The electron effective mass and 
charge are $m$ and $e$ respectively.
The screening in the electron-electron is given by the factor\cite{mar}
\begin{displaymath}
  S_\subtext{ee} \,=\,\, 
      \left (
1 + \frac{ N }{ N_{ee} } 
      \right )^{-1}
\end{displaymath}
and becomes important when the carrier 
concentration $N_\subtext{max}$ reaches a
critical value 
\begin{displaymath}
N_\subtext{ee} =
\frac{4 m {\cal E}_\infty ( k_\subtext{B} T_\subtext{e} )^2}
{\pi^2 \hbar^2 e^2}
\end{displaymath}

The frequencies associated with the collision mechanisms are in general 
dependent upon the energy. Therefore, $\nu$ is a function of $\varepsilon$ but
a smooth one\cite{mar2001}. The analysis can be considerably simplified if
we average it over the conduction band. The factor $A$ given by
\begin{displaymath}
  A \,=\,\, \frac{1}{\varepsilon_\subtext{max}-1} 
	\int_{1}^{\varepsilon_\subtext{max}} 
	d\varepsilon \frac{1}{ \sqrt{ \varepsilon } }  
	\ln{ \left [
            	\frac{ 1 + \sqrt{ 1 - 
			\frac{\D 1 }{ \D \varepsilon } } }
                   { 1 - \sqrt{ 1 -
			\frac{ \D 1}{ \D \varepsilon } } }
            \right ] }
\end{displaymath}
is this averaging process and $\varepsilon_\subtext{max}$ is the 
maximum
energy in the conduction band that can be reached by an electron\cite{mar}.

The expression for $\nu$, Eq.\ (\ref{nu}),
is one of the simplest 
approximations we can make which allows us to determine 
in an easy way the relative importance of the 
electron-electron interaction against the electron-LO phonon interaction
in terms
of the carrier concentration and 
electronic temperature. 

The electron-electron interaction which gives the nonlinear character
of the rate equation needs some clarification. We follow the ideas
of Takenaka {\em et al}.\cite{takenaka} 
and Collet {\em et al}.\cite{collet} and use static RPA (Random Phase
Approximation) to obtain\cite{mar2001,mar} 
\begin{equation}\label{eq_16}
  ZN_\subtext{max} \,\,=\,\,
    \frac{ e^2 
	\sqrt{ \pi m k_\subtext{B} T_\subtext{e} } }
         { 2^2 \hbar^2 {\cal E}_\infty } S_\subtext{ee}
\quad .
\end{equation}
The use of static RPA is justified for experiments which take place
on longer time scales\cite{haug}. In particular, for bulk GaAs, under
low temperature cw photoexcitation which creates a not very high
carrier concentrations (less than $1 \times 10^{18}$ cm$^{-3}$,
where many body and occupation effects are not important and the
electron-LO phonon and electron-electron dominates the dynamics 
of the carriers) the equations of the model are expected to be
valid\cite{mar,shah1999}. Degeneracy effects are also negligible
for concentrations below $1 \times 10^{18}$ cm$^{-3}$\cite{shah1999}.
As we mention before the equations of the model came from a
more general theory published in Ref.\cite{mar}. These general 
equations take into account degeneracy effects and can be 
generalized to take into account quantum effects, like 
exchange, by using more appropiate scattering frequencies\cite{maksym}.
However, quantum effects are expected to be important
in very short times and very small distances\cite{haug,shah1999}.

\section{Stationary states and stability analysis}
In Fig.\ 1 we show the low temperature dependence of $\chi_0^s$ as a function 
of the control parameters $\varepsilon_p$ and $\nu$.
Fig.\ 1a shows the dependence of $\chi_0^s$ upon the control parameter $\nu$
for various values of $\varepsilon_p$. We have taken 
$\omega=0.02$ for the recombination. 
When $\varepsilon_p \stackrel{<}{\sim} 16$ and the efficiency of the 
electron-electron is increased ($\nu$ is diminished)
we found a bifurcation, a range of values of $\nu$ where we have more than 
two
solutions for $\chi_0^s$.
\begin{figure}
\includegraphics[width=3.5in]{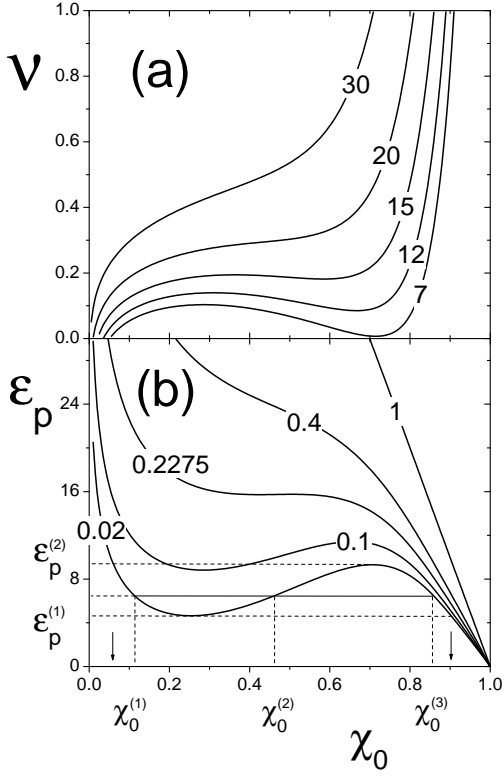}
\caption{$\chi_o^s$ as a function of the control parameters
$\nu$ (a) and $\varepsilon_p$ (b).
The number on each curve is the value of $\varepsilon_p$ (a), $\nu$ (b).
}\label{phasetr_fig_01}
\end{figure}

The dependence of $\chi_o^s$ upon the control parameter $\varepsilon_p$ 
is shown in Fig.\ 1b 
for several values of $\nu$. 
Notice that, for example, the curve with $\nu=0.02$,
for $\varepsilon_p^{(1)} < \varepsilon_p < \varepsilon_p^{(2)}$ has three
possible values of $\chi_0^s$, that are labelled $\chi_0^{(1)}$, 
$\chi_0^{(2)}$, and $\chi_0^{(3)}$.
We will show that two of them are stable and the other
one is unstable. In addition, for $\nu_c\approx 0.2275$ we have the critical
curve which separates curves which has regions of $\varepsilon_p$ with three
possible solutions for $\chi_0^s$ from curves with just one solution. 
The critical point is given by $\varepsilon_p^c$, $\chi_0^c$ and $\nu_c$
that satisfy
\begin{eqnarray}
\left( \frac{\partial \varepsilon_p}{\partial \chi_o} \right)_{\nu_c} &=& 0 
\label{eq_9}\\
\left( \frac{\partial^2 \varepsilon_p}{\partial \chi_o^2} \right)_{\nu_c} &=& 0
\end{eqnarray}
We
must point out that $\nu = \nu_o^+/ZN_\subtext{max}$, then Fig.\ 1b 
can be seen as a family of curves in which the
electron-electron interaction is changed, for example, by changing the
electron concentration. For GaAs bulk semiconductor, at 4 K, using 
$N=4.1\times 10^{17}$ cm$^{-3}$ and $T_\subtext{e}=300$ K in Eq.\ (\ref{nu})
we obtain $\nu=0.02$. The critical curve is obtained for 
$N=N_c\approx 6.11\times 10^{16}$ cm$^{-3}$.

The extrema of the curve $\varepsilon_p(\chi_0^s)$ for a given $\nu$
are $\varepsilon_p^{(1)}$ and $\varepsilon_p^{(2)}$ respectively. They 
are found by solving the equation 
$(\partial \varepsilon_p / \partial \chi_0 )_{\nu} = 0$. 
They are the boundaries of 
what is call the hysteresis region. 
If we increase $\varepsilon_p$ from a value less than $\varepsilon_p^{(1)}$,
we will move until the maximum value $\varepsilon_p^{(2)}$ is reached. Then
an abrupt jump will take place from a $\chi_0^s$ value at the maximum 
$\varepsilon_p^{(2)}$ to the value marked with the arrow at the left of
Fig.\ 1b. As we decrease $\varepsilon_p$ from a value greater than
$\varepsilon_p^{(2)}$ we will move until we reach the value
of $\chi_0^s$ at the minimum. Then the system will undergo an abrupt jump
to the value of $\chi_0^s$ marked with the right arrow in Fig.\ 1b.
In Fig.\ 2 we show this hysteresis
region for two values of the recombination parameter $\omega$.
\begin{figure}
\includegraphics[width=3.5in]{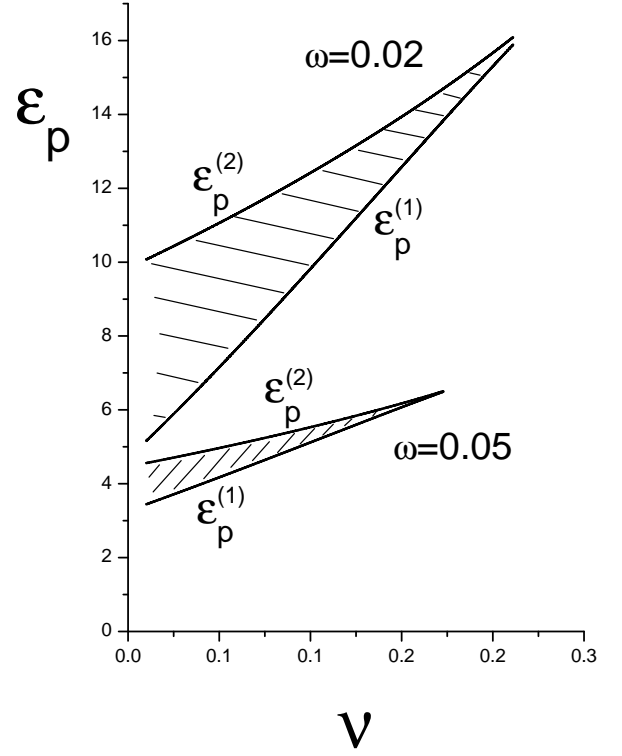}
\caption{Hysteresis region (hatched) in the phase space ($\nu$,$\varepsilon_p$)
for two values of $\omega$.}\label{phasetr_fig_02}
\end{figure}

Let us consider the perturbation to the steady state, 
$\chi(\varepsilon, \tau) = \chi^{s}(\varepsilon) + \delta\chi(\varepsilon,\tau)$ 
produced by a
time dependent excitation, $\chi_p(t)=\chi_p^s + \delta\chi_p(t)$, where 
$\chi_p^s$ is the constant excitation that produces the steady state. Notice
that $\chi_\subtext{tot}$ and $\chi_0$ are time dependent and they are given
by the relations 
\begin{eqnarray}
\chi_\subtext{tot} &=& 1 + \delta\chi_\subtext{tot}  \nonumber \\
\chi_0 &=& \chi_0^{s} + \delta\chi_0
\end{eqnarray}
From Eq.\ (\ref{phasetr_eq_02}) for $\chi_0$ and assuming that 
$\varepsilon_p > 1$, we have for its steady state 
\begin{equation}  \label{phasetr_eq_12}
\nu\chi_1^s + \chi_1^s - \chi_0^s + \chi_0^{s2}
-\omega\chi_0^s = 0
\end{equation}
and the time evolution of its perturbation $\delta\chi_0$ is given by 
\begin{eqnarray}
\frac{d}{d\tau}\delta\chi_0 &=& \nu\delta\chi_1  +
(\delta\chi_1- \delta\chi_0)(1+\delta\chi_\subtext{tot})
+ \delta\chi_\subtext{tot}(\chi_1^s - \chi_0^s) 
\nonumber \\
&& + 2\chi_0\delta\chi_0 +
(\delta\chi_0)^2 - \omega\delta\chi_0
\end{eqnarray}
which reduces to first order 
\begin{eqnarray}
\frac{d}{d\tau}\delta\chi_0 &=& \nu\delta\chi_1  +
\delta\chi_1 - \delta\chi_0 + 2\chi_0\delta\chi_0 - \omega\delta\chi_0 
\nonumber \\
&=& \left[ \nu\frac{\delta\chi_1}{\delta\chi_0} + \frac{\delta\chi_1}{
\delta\chi_0} - 1 + 2\chi_0^s -\omega \right] \delta\chi_0
\end{eqnarray}

We now define the function $\psi(\chi_0)$ that allows us to perform the
stability analysis. From Eq.\ (\ref{phasetr_eq_12}) 
\begin{equation}  \label{phasetr_eq_15}
\psi(\chi_0) \equiv \nu\chi_1 + \chi_1 - \chi_0 + \chi_0^2
-\omega\chi_0
\end{equation}
and the steady state condition (Eq.\ (\ref{phasetr_eq_12})) becomes 
\begin{equation}
\psi(\chi_0^s) = 0
\end{equation}
The derivative of $\psi(\chi_0)$ with respect of $\chi_0$ is 
\begin{equation}
\frac{d\psi(\chi_0)}{d\chi_0} = \nu\frac{d\chi_1}{d\chi_0} 
+\frac{d\chi_1}{d\chi_0} - 1 + 2\chi_0 - \omega
\end{equation}
Now, in steady state $\chi_1^s$ is function of $\chi_0^s$ and the
perturbation $\delta\chi_1$ is written as 
\[
\delta\chi_1 = \left( \frac{d\chi_1}{d\chi_0} \right)_{\chi_0=\chi_0^s}
\delta\chi_0 
\]
Therefore, 
\begin{equation}
\frac{d}{d\tau}\delta\chi_0 = \left. \frac{d\psi}{d\chi_0}
\right|_{\chi_0=\chi_0^s} \delta\chi_0
\end{equation}
and the solution for $\delta\chi_0$ is 
\begin{equation}
\delta\chi_0(\tau) = \delta\chi_0(0)\exp\left[\left. \frac{d\psi}{d\chi_0}
\right|_{\chi_0=\chi_0^s} \tau\right]
\end{equation}
A steady state, given by a $\chi_0^s$, is stable if 
\begin{equation}  \label{phasetr_eq_18}
\left. \frac{d\psi}{d\chi_0}\right|_{\chi_0=\chi_0^s} < 0
\end{equation}

Eq.\ (\ref{phasetr_eq_08}) can be written as 
\begin{equation}  \label{phasetr_eq_20}
-\frac{1}{\alpha}\ln\left[(1-\chi_0^{s})\chi_0^{s} \frac{\alpha}{\omega} 
\frac{\alpha+\beta}{e^{\alpha}-e^{-\beta}}\right]-\varepsilon_p = 0
\end{equation}
In steady state Eqs.\ (\ref{phasetr_eq_15}) and (\ref{phasetr_eq_20}) are
equivalent, therefore 
\begin{equation}  \label{phasetr_eq_21}
\psi(\chi_0)=-\frac{1}{\alpha}\ln\left[(1-\chi_0)\chi_0 \frac{\alpha}{\omega}
\frac{\alpha+\beta}{e^{\alpha}-e^{-\beta}}\right]-\varepsilon_p
\end{equation}
In Fig.\ 1b we plotted $\varepsilon_p$ as a function of $\chi_0^s$ which 
means that we plotted the function $\psi(\chi_0)$ and this
establishes the stability of the states $\chi_0^{(1)}$, $\chi_0^{(2)}$, and $
\chi_0^{(3)}$. The middle root is an unstable steady state, while $
\chi_0^{(1)}$ and $\chi_0^{(3)}$ are stable steady states.

\section{First order nonequilibrium phase transition and Maxwell construction}

A thermodynamic system remains homogeneous and stable if the criteria of
instrinsic stability is satisfied\cite{stanley}, 
\begin{equation}
\left( \frac{\partial P}{\partial V} \right)_T < 0
\end{equation}
where $P$ is the pressure of the system, $V$ is its volume
and $T$ is its temperature. When this
condition is violated a phase transition ocurrs\cite{stanley}.

Then, from Eq.\ (\ref{phasetr_eq_18}), we have the following correspondence
\cite{note01} 
\begin{eqnarray*}
\chi_0 &\rightarrow& V \\
\varepsilon_p &\rightarrow& P \\
\nu &\rightarrow& T
\end{eqnarray*}
We see that Eq.\ (
\ref{phasetr_eq_20}) turns out to be the equation of state. The homogeneous
states $\chi_0^{(1)}$ and $\chi_0^{(3)}$ are nonequilibrium stable steady
states of the system. Therefore, we call the transition between states 
$\chi_0^{(1)}$ and $\chi_0^{(3)}$, a first order out of equilibrium phase
transition. The order parameter is $\chi_0^{(3)}-\chi_0^{(1)}$ and still
remains unknown.

So far, we have assumed that the populations $\chi_0$, $\chi_1$, and so on,
are constants and homogeneous in space. Now, we assume that there exist
spatial inhomogeneities, $\chi_0 = \chi_0({\bf r})$, which produce spatial
gradients, and diffusion of this population. We also assume, from the
structure of Eq.\ (\ref{phasetr_eq_03}), that the spatial and temporal
behavior of the populations $\chi_i$, $i>0$, is given through $\chi_0$.
Then, we have the following equation 
\begin{equation}
\frac{\partial\chi_0}{\partial \tau} = \psi(\chi_0) + 
\kappa\frac{\partial^2\chi_0}{\partial z^2}
\end{equation}
We suppose, for simplicity, that $\chi_0$ depends only on the spatial
coordinate $z$. Here, $\kappa$ is the $\chi_0$ difussion constant.

Let us introduce the ``potential'' $\Phi(\chi_0)$ with the 
definition\cite{pimpale,schlogl} 
\begin{equation}  \label{phasetr_eq_23}
\psi(\chi_0) = \frac{\partial}{\partial\chi_0}\Phi(\chi_0)
\end{equation}
Then, the steady state satisfies the equation 
\begin{equation}
\kappa\frac{\partial^2\chi_0}{\partial z^2} = 
-\frac{\partial}{\partial\chi_0}\Phi(\chi_0)
\end{equation}
We have seen from Fig.\ 1b that, when $\nu=0.02$ and 
$\varepsilon_p$ is in the energy interval, 
$\varepsilon_p^{(1)}<\varepsilon_p<\varepsilon_p^{(2)}$, the system has two 
homogeneous stable
steady states. Let us find a solution $\chi_0(z)$ such that, 
$\chi_0(+\infty)=\chi_0^{(1)}$ and $\chi_0(-\infty)=\chi_0^{(3)}$. In such a
case two steady states coexist. Obviously, the ``potential'' $\Phi(\chi_0)$
has two maxima in $\chi_0^{(1)}$ and $\chi_0^{(3)}$. Coexistence ocurrs for
a value of $\varepsilon_p$ such that the two maxima are indistinguishable
for the system, 
\begin{equation}
\Phi(\chi_0^{(1)}) = \Phi(\chi_0^{(3)})
\end{equation}
Then 
\begin{eqnarray}  \label{phasetr_eq_27}
0 &=& \Phi(\chi_0^{(3)}) - \Phi(\chi_0^{(1)}) =
\int_{\chi_0^{(1)}}^{\chi_0^{(3)}} d\chi_0\psi(\chi_0)  \nonumber \\
&=& \int_{\chi_0^{(1)}}^{\chi_0^{(3)}} d\chi_0\; \left( \nu
\chi_1+\chi_1-\chi_0+\chi_0^2 - \omega\chi_0\right)
\end{eqnarray}
The last equation is the Maxwell construction for the vapor pressure in the
van der Waals gas from which the order parameter $\chi_0^{(3)}-\chi_0^{(1)}$
can be calculated. We also notice that $-\Phi$ corresponds to the Hemholtz
free potential. Moreover, the condition that $\Phi$ is at a maximum in a
stable steady state corresponds to that the generalized free potential 
$H=-\Phi$ is at a minimum. This is consistent with the condition that an
equilibrium thermodynamic system is in a state of minimum Hemholtz free
energy\cite{stanley}.

\section{Conclusions}

We have found carrier concentration bistabilities in a
low temperature nonequilibrium electron gas and we study them in 
terms of the hysteresis region. The diffusion model consider here allows
the coexistence of two phases of the open far from equilibrium gas.
The phases are distinguished by the two values of carrier concentration
at the bottom of the conduction band, namely, $\chi_0^{(1)}$ and
$\chi_0^{(3)}$. These 
two corresponding homogeneous stable steady states
are obtained using the Maxwell construction.
We introduced the function $\psi$ and the stability
condition for these states was given by Eq.\ (\ref{phasetr_eq_18}). A
generalized free potential was defined by Eq.\ (\ref{phasetr_eq_23}) from
which we obtained Eq.\ (\ref{phasetr_eq_27}) that corresponds to the vapor
pressure Maxwell construction of a van der Waals gas. 
The difference of the electron population in the bottom of the conduction
band is the order parameter and is then calculated and goes to zero when $\nu$ approaches the critical
value $\nu_c\approx 0.2275$. This is the reason we call this phase
transition a first order nonequilibrium phase transition. 
For bulk GaAs at 4 K the critical curve is obtained using
$N=N_\subtext{c}\approx 6.11\times 10^{16}$ cm$^{-3}$ and 
$T_\subtext{e}=300$ K in Eq.\ (\ref{nu}).
We have to point
out that the electron-electron interaction, which gives the nonlinear
character of the rate equations, is the necessary main ingredient for the
existence of the phase transition.


\begin{thebibliography}{99}
\bibitem{note01} A discussion on the corresponding thermodynamical variables
is made in Ref.\ \cite{landsberg76}.
\end{thebibliography}
\end{document}